\documentclass[prl,twocolumn,superscriptaddress,showpacs]{revtex4}
\usepackage{subfigure}
\usepackage{textcomp}
\usepackage{graphicx}
\usepackage{amssymb}
\usepackage{amsmath}
\usepackage{bm}
\usepackage{amsmath, amsthm, amssymb}

\begin{document}
\title{In-plane transport and enhanced thermoelectric performance in thin films of the topological insulators Bi$_2$Te$_3$ and Bi$_2$Se$_3$}


\author{Pouyan~Ghaemi}
\affiliation{Department of Physics, University of California at Berkeley, Berkeley, CA 94720}
\affiliation{Materials Sciences Division, Lawrence Berkeley National Laboratory, Berkeley, CA 94720}

\author{Roger S.~K. Mong}
\affiliation{Department of Physics, University of California at Berkeley, Berkeley, CA 94720}

\author{J.~E.~Moore}
\affiliation{Department of Physics, University of California at Berkeley, Berkeley, CA 94720}
\affiliation{Materials Sciences Division, Lawrence Berkeley National Laboratory, Berkeley, CA 94720}

\begin{abstract}
Several small-bandgap semiconductors are now known to have protected metallic surface states as a consequence of the topology of the bulk electron wavefunctions.
The known ``topological insulators'' with this behavior include the important thermoelectric materials Bi$_2$Te$_3$ and Bi$_2$Se$_3$, whose surfaces are observed in photoemission experiments to have an unusual electronic structure with a single Dirac cone~\cite{bi2se3,chen}.
We study in-plane (i.e., horizontal) transport in thin films made of these materials.   The surface states from top and bottom surfaces hybridize, and conventional diffusive transport predicts that the tunable hybridization-induced band gap leads to increased thermoelectric performance at low temperatures.  Beyond simple diffusive transport, the conductivity shows a crossover from the spin-orbit induced anti-localization at a single surface to ordinary localization.
\end{abstract}
\maketitle

\newcommand{\SqMxTwo}[4]{ \left( \begin{array}{cc} #1 & #2 \\ #3 & #4 \end{array} \right) }	
\newcommand{\SmallSqMxTwo}[4]{ \left( \begin{smallmatrix} #1 & #2 \\ #3 & #4 \end{smallmatrix} \right) }	
\newcommand{\FourVec}[4]{ \renewcommand{\arraystretch}{0.8} \left( \begin{array}{c} #1 \\ #2 \\ #3 \\ #4 \end{array} \right) }	
\newcommand{\bra}[1]{ {\langle #1 |} }
\newcommand{\ket}[1]{ {| #1 \rangle} }
\newcommand{\braket}[2]{ {\langle #1 | #2 \rangle} }
\newcommand{\vp}{ \mathbf{p} }
\newcommand{\meV}{ {\,\mathrm{meV}} }
\newcommand{\nm}{ {\,\mathrm{nm}} }
\newcommand{\K}{ {\,\mathrm{K}} }



Manipulation of phonon thermal conductivity and electronic structure by quantum confinement~\cite{hicksdresselhaus} have both been active areas of thermoelectric research, but the recent discovery that some of the best thermoelectrics are three-dimensional ``topological insulators''(TIs)~\cite{fu&kane&mele-2007,moore&balents-2006,hsieh} suggests new directions for this field.
For straightforward reasons, similar materials are required for TI behavior and for high thermoelectric figure of merit (FM)  $zT=\frac{S^2 \sigma T}{\kappa}$, where $S$ is thermopower (Seebeck) coefficient, $\sigma$ is the electrical conductivity and $\kappa$ is the thermal conductivity.
TIs require heavy elements in order to generate large spin-orbit coupling, which drives the formation of the topological surface state, and small bandgap so that the spin-orbit coupling can modify the band structure by ``inverting'' one band.

Similarly, thermoelectric semiconductors typically consist of heavy elements in order to obtain low phonon thermal conductivity and have a small bandgap (of order 5-10 times $k_B T$, where $T$ is the operating temperature) in order to obtain a large electronic power factor $\sigma S^2$.
An example is Bi$_2$Te$_3$, which has one of the highest known bulk thermopower FMs\cite{qw} and for decades has been used for thermoelectric devices\cite{old}.  
 
Many approaches have already been tried to improve thermoelectric performance of Bi$_2$Te$_3$, e.g., tuning carrier concentration (thereby increasing $S$), decreasing the thermal conductivity through alloying\cite{bk,holy}, inducing crystal defects or nanostructures\cite{dressel}, and making thin films to control transport of phonons and electrons\cite{tfilm}.  Recent photoemission experiments reveal that Bi$_2$Te$_3$ is a TI with  a single Dirac cone on the surface\cite{chen}, consistent with electronic structure predictions\cite{zhangn}.
It is natural to ask whether this new surface state could lead to improved thermoelectrics; this question is a major motivation for our study.

We start by deriving the effective surface Hamiltonian on symmetry grounds, then use the resulting band structure to compute the thermoelectric properties using standard diffusive transport.  Because of the large bulk gap ($\Delta_\textrm{gap} \sim 165\meV$), the chemical potential could be tuned to be in the bulk band gap\cite{chen}, and in this regime we can study the transport properties of the surface states independently.  In closing, anti-localization effects in a thin film resulting from the spin-orbit-induced Berry phase are discussed.  

The physical system to be studied here is a thin film of Bi$_2$Te$_3$.
If the film is thin enough the surface states on both sides hybridize and open a gap\cite{gap1,gap2,gapp}. Although the film thickness required to open an observable gap for TI surface states ($\sim 1\mbox{-}10 \nm$) is small, it is accessible with current growth techniques and very recently the hybridization gap has been observed by in-situ photoemission on thin films of Bi$_2$Se$_3$.~\cite{gapexp}


We obtain the effective Hamiltonian of a thin film in the basis: $\ket{u\!\uparrow}$, $\ket{u\!\downarrow}$, $\ket{d\!\uparrow}$, $\ket{d\!\downarrow}$. The basis states are denoted by spin and the surface (top($u$)/bottom($d$)) at which they are localized.  The general effective Hamiltonian takes the block form:
\begin{align}
	H(\vp) = \SqMxTwo{ H_{u} }{ V_{du} }{ V_{ud} }{ H_{d} }
	\label{GeneralEffSurfaceHamiltonian}
\end{align}
$H_{u}$ and $H_{d}$ are hermitian, while $(V_{du})^\dag = V_{ud}$.
The diagonal block terms $H_{u}$ and $H_{d}$ are the effective Hamiltonians in the semi-infinite slab.
The off-diagonal terms $V_{du}$, $V_{ud}$ describe the coupling of the two surfaces.

The parity (inversion) symmetry in the Bi$_2$Te$_3$ structure puts constraints on the form of the Hamiltonian in (Eq.\ref{GeneralEffSurfaceHamiltonian}) that are strongest if the top and bottom surfaces are related by parity.  Parity ($\Pi$) exchanges the top and bottom surfaces, changes the sign of the momenta $\vp$, but has no effect on spin: $\Pi H(\vp) \Pi^{-1} = H(-\vp)$.
Time reversal ($\mathcal{T}$) exchanges the spins and flip the momenta: $\mathcal{T} H(\vp) \mathcal{T}^{-1} = H(-\vp)$.
Explicitly $\Pi$ and $\mathcal{T}$ are
\begin{align*}
		\Pi & = \SmallSqMxTwo{}{\;1\;}{\;1\;}{}
	&	\mathcal{T} & = -i \SmallSqMxTwo{\sigma^y}{}{}{\sigma^y} \mathcal{K},
\end{align*}
where $\mathcal{K}$ is the complex conjugation operator.

The combination of parity and time reversal implies $\Pi \mathcal{T} H(\vp) (\Pi\mathcal{T})^{-1} = H(\vp)$, which
restricts the form of the effective Hamiltonian via $\sigma^y H_{u}^\ast \sigma^y = \sigma^y H_{u}^T \sigma^y = H_{d}$ and $\sigma^y V_{du}^T \sigma^y = V_{du}$.
It useful note the identities $\sigma^y (\sigma^i)^T \sigma^y = -\sigma^i$ for $i=x,y,z$ and $\sigma^y I^T \sigma^y = I$.
If we decompose the matrix $H_{u} = cI + \mathbf{a}\cdot\bm{\sigma}$, then $\sigma^y H_{u}^T \sigma^y$ flips the sign of the Pauli matrices components $\mathbf{a}\cdot\bm{\sigma}$ but leave the identity component $cI$: $H_{d} = cI - \mathbf{a}\cdot\bm{\sigma}$. Since $H_{u}$ is hermitian, $c$ is real and $\mathbf{a}$ is a real vector.
A similar argument with $V_{du} = \Delta_f I + \bm\Delta_f\cdot\bm\sigma$ will show that the Pauli matrices components must vanish, and that $V_{du} = \Delta_f I$ is a multiple of identity, $\Delta_f$ is a complex number.
\begin{align}
	H(\vp) = \SqMxTwo{\mathbf{a}(\vp)\cdot\bm{\sigma}} {\Delta_f(\vp)}
			{\Delta_f^\ast(\vp)} {-\mathbf{a}(\vp)\cdot\bm{\sigma}} + c(\vp)
\end{align}
Without loss of generality, we can find a suitable gauge transformation such that $\Delta_f$ is real; this gauge transformation is allowed by the phase ambiguity in the parity operator between the top and bottom states.

Using parity to relate the parameters $\mathbf{a}, \Delta_f, c$ at opposite momenta, there are further constraints: $\mathbf{a}(-\vp) = -\mathbf{a}(\vp)$, $\Delta_f(-\vp) = \Delta_f(\vp)^\ast$ and $c(-\vp) = c(\vp)$.
Near the Dirac cone, we expand the Hamiltonian to the first order in $\vp$. As $\Delta_f$ and $c$ are even in $\vp$, they can be treated as constant,
while $\mathbf{a}$ is odd in $\vp$ and we only keep the linear term, of the form $V_D \vp\cdot\bm\sigma$ for Bi$_2$Te$_3$ (isotropic velocity).

	The resulting Hamiltonian is
		$H = \SmallSqMxTwo{H_u} {V_{du}}{V_{du}^\dagger} {H_d}$,
	where $H_u$($H_d$) corresponds to the upper (lower) surface of the film when the two surface wavefunctions have no overlap (thick film), and $V_{du}$ corresponds to the hybridization between the two surfaces.
From first-principles electronic structure calculations\cite{zhangn} one can find parameters for $H$ and the band structure of the surface modes:
\begin{eqnarray}
H(\vp) &=& \SqMxTwo{V_D\vp\cdot\bm{\sigma}} {\Delta_f}
			{\Delta_f} {-V_D\vp\cdot\bm{\sigma}} \label{hamil} \\
			E_\vp&=&\pm\sqrt{(V_D\vp)^2+\Delta_f^2}\label{dispers},
\end{eqnarray}
$V_D \approx 4.05\times 10^{5}\ m/s$\cite{chen} (in thin films is between $3.4\times 10^{5}$ and $4.6\times 10^{5}\ m/s$ \cite{gapexp}) indicates the Dirac velocity and $\vp$ is the momentum with respect to Dirac point.

\begin{figure}
\includegraphics[width=5cm]{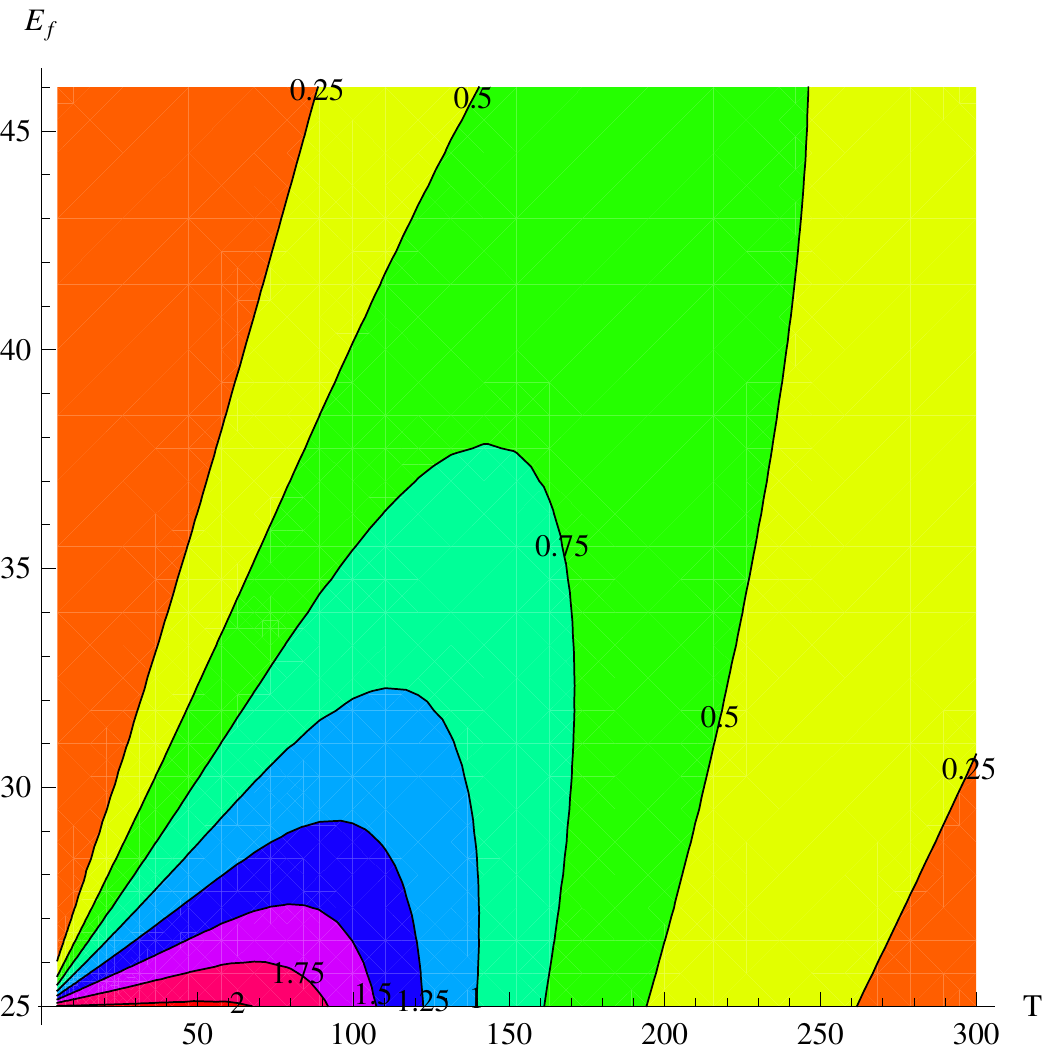}
\caption{FM of surface states alone as a function of chemical potential and temperature.  At low temperature the increase in the attainable FM becomes dramatic, but maximizing this increase requires good chemical potential control.}
\label{sfz}
\end{figure}

Now using the electron dispersion in Eq.\ref{dispers} we obtain the in-plane transport coefficients for the surface states\cite{mahan}:
\begin{eqnarray}
\sigma &=& e^2\int_{-\infty}^{\infty} d\varepsilon \frac{\partial f\left(\varepsilon \right) }{\partial \varepsilon } \Sigma(\varepsilon) \label{cond}	\\
T \sigma S &=& e\int_{-\infty}^{\infty} d\varepsilon \frac{\partial f\left(\varepsilon \right) }{\partial \varepsilon } \Sigma(\varepsilon)\left(\varepsilon-E_f\right)	\label{tpow}\\
T\kappa_0 &=& \int_{-\infty}^{\infty} d\varepsilon \frac{\partial f\left(\varepsilon \right) }{\partial \varepsilon } \Sigma(\varepsilon) \left(\varepsilon-E_f\right)^2 \label{tcond}
\end{eqnarray} 

where as before $\sigma$ is conductivity and $S$ Seebeck coefficient, but $\kappa_0$ is the electronic thermal conductivity at constant electrochemical potential, which is related to the thermal conductivity for zero electric current through $\kappa_e=\kappa_0-S^2\sigma T$.
$\Sigma\left(\varepsilon\right)=N(\varepsilon)\tau(\varepsilon)\nu_x(\varepsilon)^2$ is the conductivity density and $f(\varepsilon)$ is the Fermi-Dirac distribution function.
Experiments on bulk Bi$_2$Te$_3$ and its topologically protected surface states allow estimation of the hybridization gap\cite{nolas,chen} which we assume $\sim 25 \meV$.
This could be achieved with recently created films of thickness a few nanometers~\cite{gap1,gap2,gapexp}.
In Figure~\ref{sfz} we show the FM ($zT$) for the surface states alone.

It can be seen that at low temperatures (i.e. $150\K$ and below) FM from surface states is large compared with current low temperature thermoelectric materials (e.g. Na$_x$CoO$_2$\cite{coo} and CsBi$_4$Te$_6$\cite{cs}).
In the thin film, transport properties of the surface states should be considered in parallel with the bulk of the thin film.
Because of its thermoelectric performance ($zT \sim 0.6$) at room temperature, the bulk properties of Bi$_2$Te$_3$ have been widely studied\cite{nolas,old}.
The FM for the whole film is
\begin{equation} \label{merit}
zT = \frac{\left(S_{f}\ \sigma_{f}+S_{b}\ \sigma_{b}\ d\right)^2\ T}{\left(\sigma_{f}+\sigma_{b}\ d\right)\left(\kappa_f+\kappa_{b}\ d\right)}.
\end{equation}
$f$ and $b$ stand for surface and bulk properties and $d$ denotes the film thickness.
Using the bulk properties\cite{kaibe,proceed} for the temperature range where surface states have high FM, we can estimate the total thermoelectric FM for the thin film (but note that using a single value for the bulk contribution at a given temperature, since the transport properties even in bulk depend strongly on Fermi level). Figure 2 shows $zT$ versus Fermi level for $1\nm$ thick films at $175\K$, $150\K$, $100\K$ and $50\K$.
The straight line in each plot indicates $zT$ for CsBi$_4$Te$_6$\cite{cs}. 

Thermoelectricity from ballistic gapless 1D edge states in a 2D quantum spin Hall (QSH) nanoribbon was discussed recently by Takahashi and Murakami~\cite{murakami}. Thin films of some 3D TIs are in QSH states~\cite{gap2}, so the edge states could contribute also in a narrow film.  However, in the QSH proposal the chemical potential is in the bulk bands; here it is in the gap.

\begin{figure}
\includegraphics[height=6cm,width=5cm,bb=80 90 270 300]{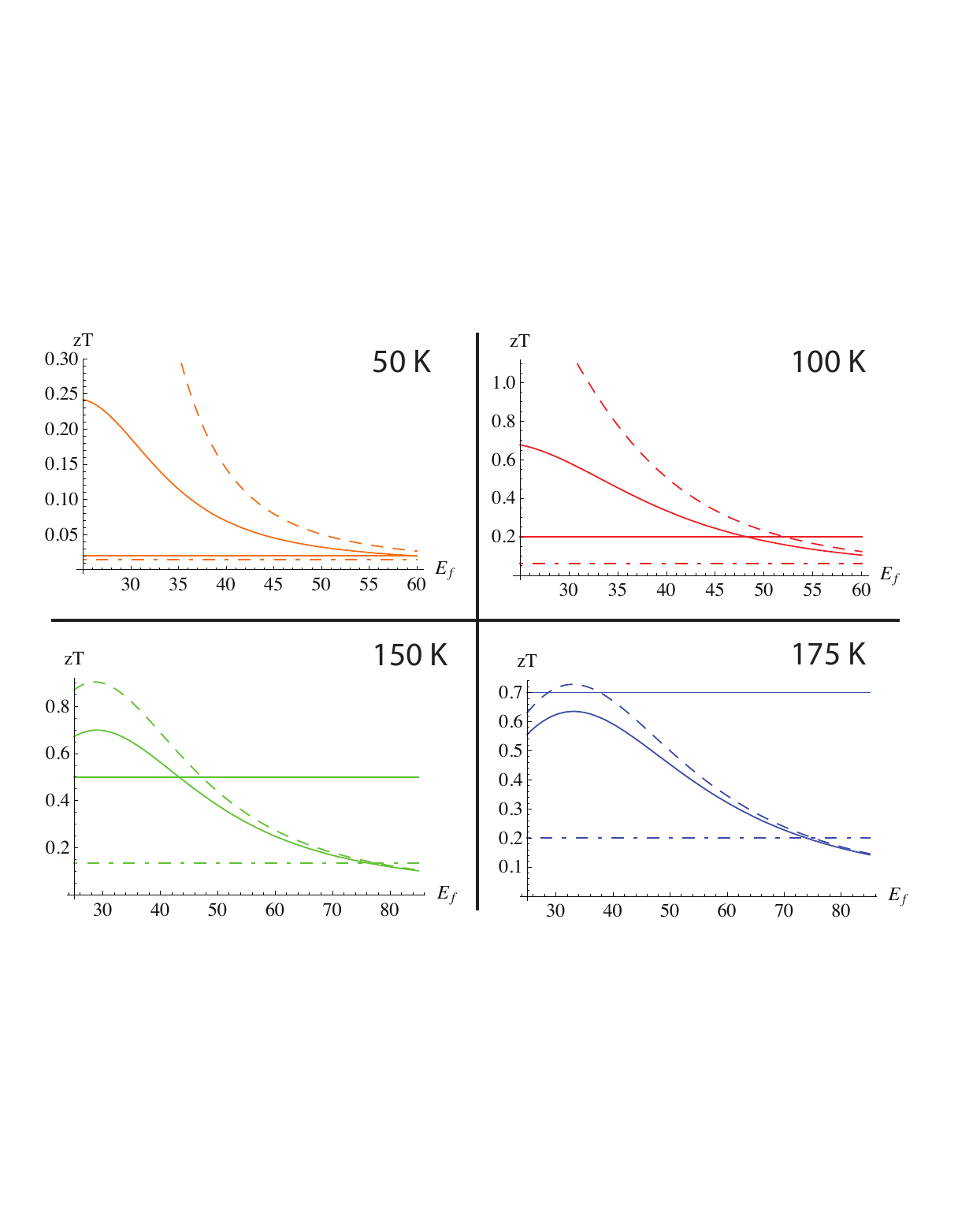}
\caption{FM for the thin film including bulk contributions at 50, 100, 150, and 175 Kelvin. The straight line in each figure corresponds to the best known FM at the same temperature. Dashed line indicates the FM for the surface states alone and Dashed-Dotted line indicates FM for bulk Bi$_2$Te$_3$. $\huge{\sigma_b} \times 10^{-5}\  (\Omega m)^{-1} =2,\ 1.25,\ 0.83,\ 0.72$, $\Huge{\kappa_b}\  (W K^{-1} m^{-1})=3.35, 2.83, 2.35, 2.2$  and $\huge{S_b} \times 10^6\  (V K^{-1})=70, 115, 150, 171$ at $T=50,100,150,175$ K. \cite{kaibe,proceed}}
\label{ztfilm}\end{figure}

As is evident from Figure \ref{ztfilm}, at temperatures below 150 K, which are important for several Peltier cooling applications, the thermoelectric performance of the TI thin film is significantly enhanced because of the high FM of the protected edge states.  At low temperature the bulk contribution is smaller than the surface contribution so that the unknown chemical potential dependence of bulk properties is not too significant.
Crucially, the gap in the hybridized surface mode band structure can be controlled by tuning the thickness of the film to get high FM in a specific temperature range.
The geometry of thin films is also very effective in reduction of phonon thermal conductivity, so there will be even larger enhancement for the TI thin films.
The same approach applies as well to Bi$_2$Se$_3$ and Bi$_2$Se$_x$Te$_{3-x}$ alloys.

\begin{figure}
\includegraphics[width=4cm]{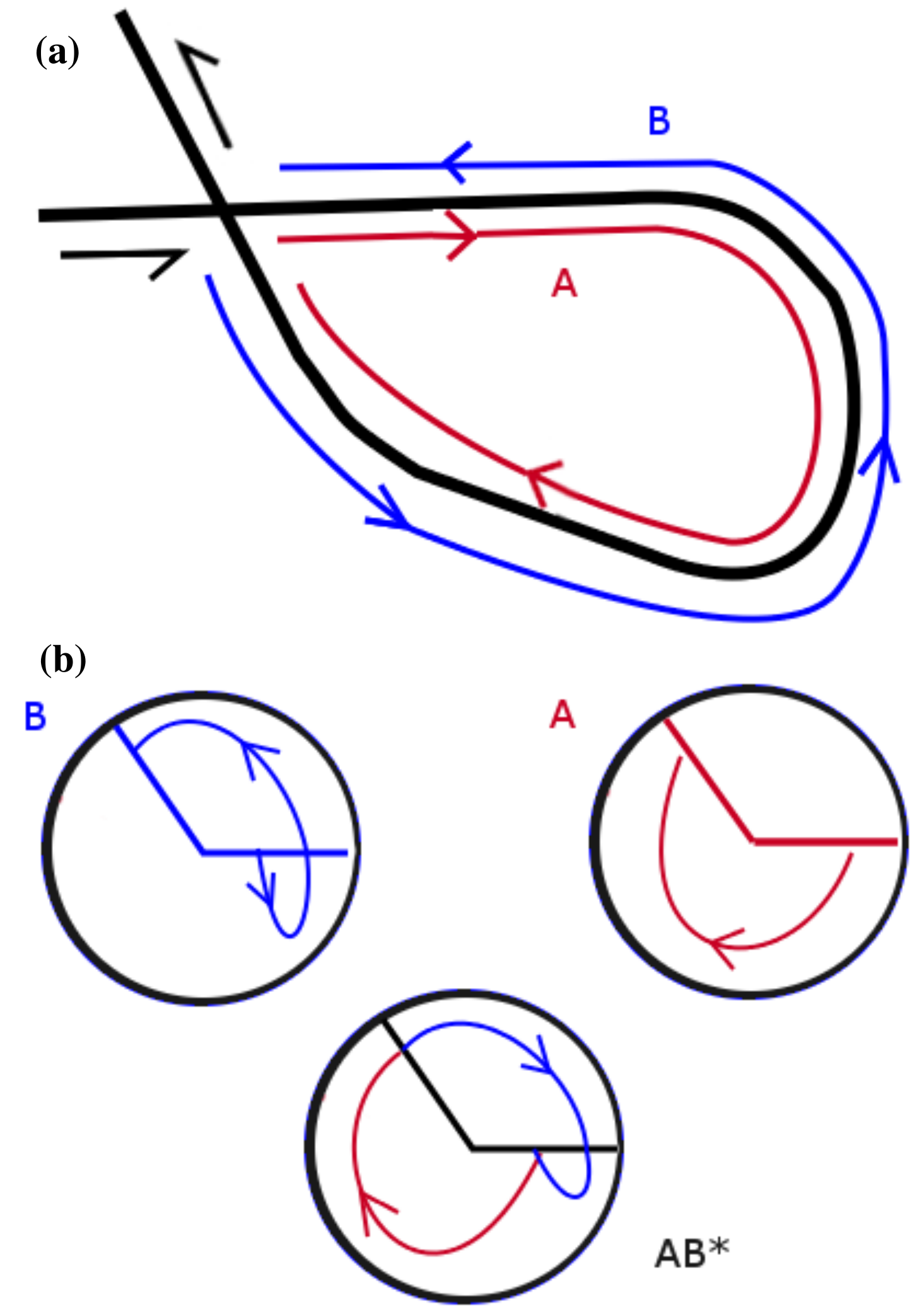}
\caption{Back-scattering process: a) Two trajectories (A and B) leading to back scattering in real space traveled in opposite directions. b) A and B trajectories in momentum space. They add up to rotation of momentum by $2\pi$.}
\label{cloop}\end{figure}

The conductivity in Eq.\ref{cond} is calculated using Boltzmann formalism \cite{prev} i.e. classical trajectories of electrons.
Quantum mechanical corrections appear as the interference of quantum phases of different trajectories.
In a simple picture, the amplitude for an electron moving between two positions ($a$ and $b$) is given by $|\mathbf{A}_{ab}|=|\sum_n A_n e^{i \phi_n}|$ where $n$ labels different classical paths between $a$ and $b$ and $\phi_n$ represents the corresponding quantum phase\cite{feynman}. This phase is usually random and averages to zero (i.e. $|\mathbf{A}_{ab}|^2=\sum_n |A_n|^2$ ), which leads to the Boltzmann conductivity as in Eq.\ref{cond}.
On the other hand, for closed trajectories where the electron returns to a region of size comparable to its quantum wavelength, quantum interference plays an important role.
For diffusive motion, as in a metal, there is a finite probability for closed paths (fig. \ref{cloop}).
Such closed loops in ordinary metals lead to reduction of the conductivity (i.e. localization)~\cite{prev}.
The fact that such paths can be taken in two opposite directions means that the relative phases do not average to zero; the corresponding quantum correction is $\frac{\delta \sigma}{\sigma} = - \cos\left(\phi^+ -\phi^-\right)$ where $\phi^+$ and $\phi^-$ corresponds to phases for two opposite trajectories.


\newcommand{\vpF}{ {\vp_{_F}} }		
Quantum corrections affect the conductivity of materials strongly at mesoscopic length scales\cite{sharvin}. Here we show that the chiral nature of states on the Fermi surface (FS) can increase the mean-free path even when hybridization is present.  At each momentum, the effective surface Hamiltonian (Eq.\ref{hamil}) has two degenerate eigenstates:
\begin{align}
\ket{a_\vp} = \frac{1}{\sqrt{2}E_\vp} \left(\begin{array}{c}
		0 \\
		\Delta_f	\\
		-p\; e^{-i\theta_\vp}	\\
		E_\vp
	\end{array}\right)
&&	\renewcommand{\arraystretch}{0.7}
\ket{b_{\vp}} = \frac{1}{\sqrt{2}E_\vp} \left(\begin{array}{c}
		E_\vp \\
		p\; e^{i\theta_\vp}	\\
		\Delta_f	\\
		0
	\end{array}\right)
\end{align}
where for simplicity $V_D=1$, $p$ and $\theta_{\vp}$ are the polar coordinates of $\vp$ such that $(p_x,p_y)\!=\!p\,(\cos\theta_\vp,\sin\theta_\vp)$.
On the FS $p$ is fixed so states are labeled by $\theta_\vp$'s. 

Although at each momentum $\textbf{p}$ states $\ket{a_\vp}$ and $\ket{b_\vp}$ are orthonormal, they are not at different momenta:
\begin{align}
\braket{a_{\vp'}}{a_\vp} &= \frac{2|\Delta_f|^2 + p^2 [1 + e^{i( {\theta_\vp}_{'}-\theta_\vp )} ]}{2 E_\vp^2}	\\
\braket{b_{\vp'}}{b_\vp} &= \frac{2|\Delta_f|^2 + p^2 [1 + e^{i( \theta_\vp - {\theta_\vp}_{'} )} ]}{2 E_\vp^2}	\\
\braket{a_{\vp'}}{b_\vp} &= \frac{p\,\Delta_f}{2 E_\vp^2}\left(e^{i\theta_\vp} - e^{i{\theta_\vp}_{'}}\right).
\end{align}
$\braket{b_\vp}{a_{\vp'}}$ is small for small angle scatterings and small $\Delta_f$. 
Any superposition of these two states (i.e. $\ket{\psi_\vp}= \alpha\ket{a_\vp} + \beta\ket{b_\vp}$) is also an eigenstate with energy $E_\vp = \sqrt{\Delta_f^2+p^2}$. 

Smooth impurities on the surface scatter different states on the FS into each other (i.e. they change $\theta_\vp$) and lead to diffusive motion of quasiparticles and classical Boltzmann conductivity\cite{prev} $\sigma=\frac{n e^2\tau}{m}$ which we used in our calculations. Quantum effects could be cast as quantum phases that quasiparticles gain while moving along different classical trajectories\cite{feynman}. Quantum corrections are due to interference of these different phases.
However the phase difference between different trajectories are generally random and on average give no contribution to the conductivity\cite{prev}, with the exception of self-intersecting trajectories. Such trajectories could be taken in two different direction. The relative phase is not random and leads to quantum interferences. 

As mentioned above a general state on the FS is $\ket{\psi_\vp}= \alpha\ket{a_\vp} + \beta\ket{b_\vp}$. But not all of these states can form a self intersecting trajectory. 
When $\Delta_f=0$ (i.e. decoupled surfaces) clearly the states $\ket{a_\vp}$ is only affected by impurities in bottom surface and $\ket{b_\vp}$ is affected by impurities in top surface. When impurities in top and bottom surfaces are independently distributed, one can consider state $\ket{a_\vp}$ or $\ket{b_\vp}$ move over a self-intersecting trajectory independently, but not coherently together. In other words a mixed state of the form $\ket{\psi_\vp}= \alpha\ket{a_\vp} + \beta\ket{b_\vp}$ with $\alpha \neq 0$ and $\beta \neq 0$ cannot go over a classical self-intersecting trajectory, since the direction of motion ($\theta_\vp$) in $\ket{a_\vp}$ and $\ket{b_\vp}$ changes independently with scatterings in bottom and top surfaces respectively. As a result quantum interference effects are important for pure $\ket{a_\vp}$ and pure $\ket{b_\vp}$ states not the mixed states.

 When $\Delta_f$ is non-zero but small, $\ket{a_\vp}$ and $\ket{b_\vp}$ are mainly (but not entirely) localized in bottom and top surfaces. An important feature of $\ket{a_\vp}$ and $\ket{b_\vp}$ is that even when $\Delta_f \neq 0$ they only depend on direction of momentum ($\theta_\vp$) in bottom and top surfaces respectively. Even though they have non-zero components on opposite surfaces, they depend on the direction of momentum only in one surface. So as in the case with $\Delta_f=0$, when the impurities are randomly distributed on
top and bottom surfaces, $\ket{a_\vp}$ and $\ket{b_\vp}$ can move over
self-intersecting trajectories independently.  However, in a mixed state,
$\ket{\psi_\vp}= \alpha\ket{a_\vp} + \beta\ket{b_\vp}$, the directions $\theta_\vp$
change independently in two surfaces and do not simultaneously form closed loops
(see fig. \ref{cloop}). As a result, in order to track the quantum interference effects we should consider $\ket{a_\vp}$ and $\ket{b_\vp}$ independently. The quantum phase of each trajectory is given by the associated Berry phase\cite{berry}:
\begin{equation}\begin{split}
\phi_a &= -i \int_0^T dt \bra{a_{\vp_{_t}}} \frac{\partial}{\partial t} \ket{a_{\vp_{_t}}}
		=  -\frac{p^2}{2E_\vp^2} \left( \theta_{\vp_{_T}} - \theta_{\vp_{_0}} \right)
		\\
\phi_b &= -i \int_0^T dt \bra{b_{\vp_{_t}}} \frac{\partial}{\partial t} \ket{b_{\vp_{_t}}}
		= \frac{p^2}{2E_\vp^2} \left( \theta_{\vp_{_T}} - \theta_{\vp_{_0}} \right)
\end{split}\end{equation}
 
\begin{figure}
\includegraphics[width=5cm]{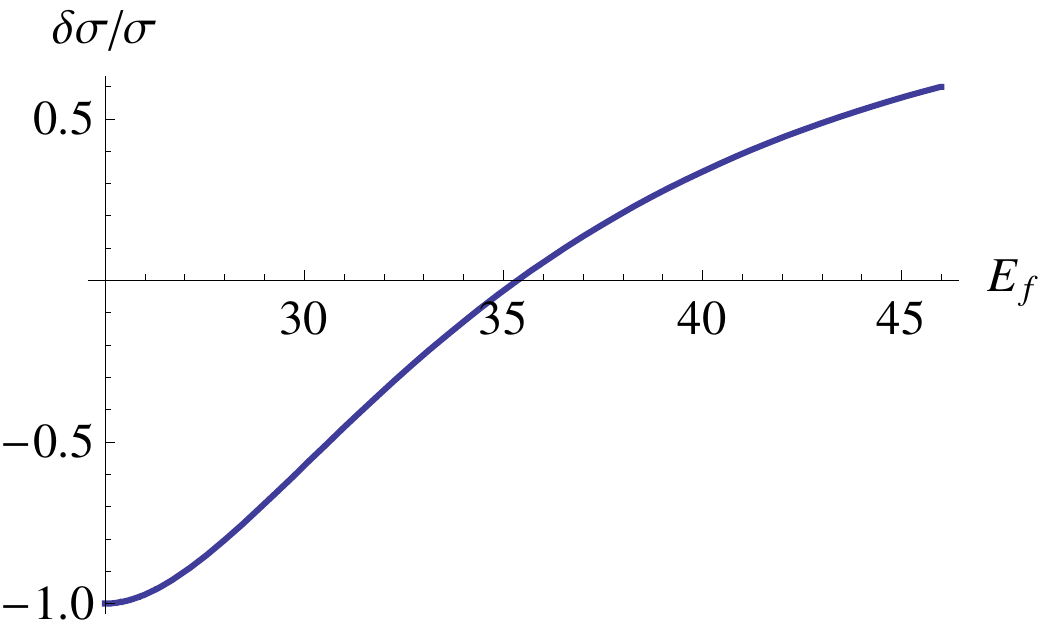}
\caption{Quantum corrections to the conductivity. Comparing
with Fig. \ref{ztfilm} we see a finite range over which antilocalization coexists
with enhanced $zT$ predicted from ordinary diffusive transport (Eq. \ref{cond},\ref{tpow},\ref{tcond}).}
\label{antc}
\end{figure}

The quantum amplitude of the closed trajectory corresponds to the sum of the of amplitudes when the self-intersecting path is taken in opposite directions; i.e. $|\left(e^{i\phi_1}+e^{i\phi_2}\right)/\sqrt{2}|^2=1+\cos\left(\phi_1-\phi_2\right)$ where $\phi_1$ and $\phi_2$ correspond to the Berry phase of the passing over the closed loop in opposite direction. As can be seen in figure \ref{cloop} $\left(\phi_1-\phi_2\right)$ corresponds to the phase associated with rotation of $\theta_p$ by $2\pi$.
The corresponding Berry phase could be calculated: $\phi_1-\phi_2 = \int_{\theta_0}^{\theta_0+2\pi} dt\, \bra{ \Psi(\vp_t) } \frac{\partial}{\partial t} \ket{ \Psi(\vp_t) } = \pm\frac{p^2}{p^2+\Delta_f^2}\pi=\pm\frac{E_\vp^2-\Delta_f^2}{E_\vp^2} \pi$ for $\Psi=\ket{a}$ or $\ket{b}$.
The two limits $p \rightarrow 0$ ($\Delta\phi \rightarrow 0$) and $p \gg \Delta_f/V_D$ ($\Delta\phi \rightarrow \pm\pi$) lead to localization and anti-localization behavior respectively.
With increasing the chemical potential, localization behavior changes into anti-localization, which is expected for a single TI surface. Comparing fig. \ref{antc} with the Seebeck results in figure \ref{sfz} we see that in some range of chemical potential where the films have high thermoelectric performance, the anti-localization effect is also present. This leads to another advantage of the TI thin films: the quantum corrections increase the conductivity.

%


We found that simply creating a nanometer-scale thin film of Bi$_2$Te$_3$ generates a hybridization-induced bandgap of the unconventional surface states that can be tuned by film thickness.
Combined with some residual robustness against impurity scattering, this leads to an increased low-temperature FM of the film.
The same technique can equally well be applied to Bi$_2$Se$_3$ or alloys in this class.  While we have concentrated on the specific case of a thin film (or single superlattice layer) in this paper, the topological surface state of these materials is an important physical feature that will also affect thermoelectric transport in other nanoscale geometries.

 The authors acknowledge conversations with K. Hippalgaonkar and R. Ramesh, and support from BES DMSE (P.~G. and J.~E.~M.)

\bibliography{topothermoprl2}


\end{document}